\documentclass[twocolumn,trackchanges,twocolappendix]{aastex631}

\usepackage{amsmath}

\begin{document}

\title{Shear-gravity transition determines the steep velocity dispersion-size relation in molecular clouds: confronting analytical formula with observations}

\author[0009-0004-5094-3522]{Yi-Heng Xie}
\affiliation{South-Western Institute for Astronomy Research, Yunnan University, Chenggong District, Kunming 650500, P. R. China}

\author[0000-0003-3144-1952]{Guang-Xing Li}
\altaffiliation{E-mail: gxli@ynu.edu.cn (G-XL)}
\affiliation{South-Western Institute for Astronomy Research, Yunnan University, Chenggong District, Kunming 650500, P. R. China}

\begin{abstract}
The velocity dispersion-size relation ($\sigma_{\rm v}\sim R^{\beta}$) is a crucial indicator of the dynamic properties of interstellar gas, where the slope is considered as $\beta \sim 0.5$. Recent observations reveal a steep velocity dispersion-size relation with the slope $\beta> 0.6$, which cannot be explained by a single mechanism with only gravity ($\beta\sim0.5$) or shear ($\beta \sim 1$). We present a two-component model $\sigma_{\rm v_{total}} = \lambda_1 \sigma_{\rm v_{g}} + \lambda_2 \sigma_{\rm v_{shear}} = A[(GM/R)^{\frac{1}{2}} + f(R/t_{\rm shear})]$ to explain the steep velocity dispersion-size relation for clouds larger than several parsecs in observations from e.g. Miville-Desch\^{e}nes et al. (2017), Zhou et al. (2022) and Sun et al. (2024). We find that, above several parsecs, the velocity dispersion of small clouds is mainly caused by self-gravity, while large clouds are primarily affected by shear, and these two regimes are linked by a gradual transition with a transition scale $\sim100$ pc---the scale height of the Galactic molecular gas disk. The variation of cloud velocity dispersion-size relation with the Galactocentric distance results from the variation of both cloud internal density structure and Galactic shear rate. Our two-component model captures how the dynamics of the molecular gas can be affected by both internal and external factors, and we expect it to be applied to data from galaxies with different physical conditions to reveal the physics. 

\end{abstract}


\section{Introduction} 
\label{sec:intro}

The dynamic behavior of interstellar gas is controlled by various factors such as gravity, turbulence, galactic shear, stellar feedback, and magnetic field. Turbulence is regarded as playing a crucial role by supporting molecular clouds against collapse \citep{Fleck1980, Larson1981}. In the 1980s, \citet{Larson1981} derived a power law velocity dispersion-size relation $\sigma_{\rm v} \sim R^{\beta}$ ($\beta=0.38$) of molecular clouds. He proposed that this relation is similar to the Kolmogorov law ($\sigma_{\rm v} \sim L^{\frac{1}{3}}$) for subsonic incompressible turbulence, where kinetic energy is injected on a large scale, transferred to the small scale, and finally dissipated. Previous studies have derived a new slope of $\beta \sim 0.5$ through different observations (e.g. \citealt{Solomon1987, McKee2007, Garcia2014, Rice2016}), which can be interpreted by virial equilibrium \citep{Solomon1987} or compressible supersonic turbulence \citep{Brunt2004}. However, recent studies obtained a larger slope of $\beta > 0.6$ (e.g. \citealt{Shetty2012, Hughes2013, Miville2017, Kauffmann2017, Liu2021, zhou2022, Colman2024, Sun2024}), which remains challenging to explain.

The physical processes that cause the scale-dependent velocity dispersion of interstellar gas remain challenging. Potential mechanisms include gravitational instabilities, galactic rotation, star feedback (e.g. stellar wind, supernovae), etc. \citep{Elmegreen2004}. The gravitational potential energy of the gas can be converted into the kinetic energy of turbulence during cloud collapse (e.g. \citealt{Wada2002, Ballesteros-Paredes2011, IBANEZ2016, Ibanez2017, Traficante2018}). Galactic shear induced by galactic rotation can lead to different Keplerian velocities of gas at different galactic locations and serve as a continuous source driving the turbulence of molecular clouds (e.g. \citealt{Fleck1981}). However, the steep velocity dispersion-size relation with slope $\sim 0.6-0.8$ cannot be explained by a single mechanism with only gravity or shear. If we assume that the velocity dispersion of molecular clouds is completely caused by self-gravity, the corresponding relation is $\sigma_{\rm v}\sim R^{0.5}$. Meanwhile, assuming that the velocity dispersion is only caused by the shear, the corresponding relation is $\sigma_{\rm v}\sim R^{1}$.

We present a two-component model, including shear and self-gravity, to explain the steep slope of the velocity dispersion-size relation in molecular clouds larger than several parsecs. We use our model to explain the observation result traced by young stellar objects (YSOs) in the solar vicinity \citep{zhou2022} and determine its full form. We further apply our model to the cloud sample within the Milky Way disk \citep{Miville2017, Sun2024} to predict the variation of the velocity dispersion-size relation with different Galactocentric distances.

\section{Steep slope and variation of velocity dispersion-size relation as challenges}

{\textbf {Puzzle 1: Steep slope of $\sigma_{\rm v}-R$ relation.}} Recent observations have obtained a steep velocity dispersion-size relation with a slope $\beta\sim0.6-0.8$ for the molecular clouds in different environments such as the Galactic Center, the Galactic disk, and extragalactic galaxies. \citet{Kauffmann2017} derived $\sigma_{\rm v} = (5.5\pm 1.0)\ (R_{\rm eff}/\rm pc)^{0.66\pm 0.18}\ (km\ s^{-1})$ in the central molecular zone (CMZ) probed by $\rm N_2H^+$ (3-2) transition. \citet{Miville2017} obtained $\sigma_{\rm v} = 0.48\ (R/\rm pc)^{0.63\pm 0.30}\ (km\ s^{-1})$ for the clouds in the entire Galactic disk using $^{12}\rm CO$ emission. \citet{Liu2021} reported $\sigma_{\rm v} = (-0.30\pm 0.17)\ (R/\rm pc)^{0.82\pm 0.13}\ (km\ s^{-1})$ of the clouds in the lenticular galaxy NGC 4429 probed by $\rm ^{12}CO\ (J=3-2)$ observation. \citet{zhou2022} obtained a two-dimensional velocity dispersion-size relation of $\sigma_{\rm v}(\rm 2D) = 0.74\ (R/\rm pc)^{0.67}\ (km\ s^{-1})$ probed by the proper motion measurement of YSO associations in the solar vicinity. \citet{Sun2024} derived $\sigma_{\rm v} = 0.26\ (R/\rm pc)^{0.65\pm 0.004}\ (km\ s^{-1})$ of the resolved clouds within the northern outer Galactic disk using $\rm ^{12}CO\ (J=1-0)$ emission line. In simulation, \citet{Colman2024} reported $\sigma_{\rm v}\propto R^{0.75}$ for large clouds. \citet{Cen2021} also proposed a new form of Larson's relation of $\sigma_{\rm R}= {\alpha_{\rm vir}}^{\frac{1}{5}} \sigma_{\rm pc} (R/ \rm 1pc)^{\frac{3}{5}}$, where $\alpha_{\rm vir}$ is the virial parameter of clouds and $\sigma_{\rm pc}$ is the velocity dispersion of turbulence at 1 pc.
    
These results reveal that such a steep velocity dispersion-size relation is not rare in the galaxies. These slopes cannot be explained by a single mechanism with only gravity ($\sigma_{\rm v}\sim (GM/R)^{\frac{1}{2}}$) or shear ($\sigma_{\rm v}\sim R/t_{\rm shear}$), where $M$ and $R$ are the mass and size of molecular clouds, $t_{\rm shear}$ is the shear timescale. According to Larson's relation \citep{Larson1981} molecular clouds follow a mass-size relation of $M\sim R^2$, implying that at the same Galactocentric distance, the self-gravity-induced and shear-induced velocity dispersion follow a scaling relation of $\sigma_{\rm v_{g}}\sim R^{0.5}$ and $\sigma_{\rm v_{shear}}\sim R^{1}$ respectively. 

{\textbf {Puzzle 2: Variation of $\sigma_{\rm v}-R$ relation with Galactocentric distance.}} The mechanism that leads to the variation of molecular cloud velocity dispersion-size relation with Galactocentric distance remains a challenge. In observations, \citet{Miville2017} found a larger normalization of velocity dispersion-size relation at the inner disk than the outer Galaxy. \citet{Sun2024} also derived a smaller normalization factor in the outer Galactic disk. 

\citet{Heyer2009} found that the normalization of velocity dispersion-size relation varies with cloud surface density as $\sim \Sigma^\frac{1}{2}$, while \citet{Miville2017} derived a similar relation of $\sigma_{\rm v}\sim 0.23(\Sigma R)^{0.43\pm 0.14}$. Based on this, self-gravity is regarded as an important energy source for observed non-thermal motions in molecular clouds (e.g. \citealt{Ballesteros-Paredes2011, IBANEZ2016, Traficante2018}). \citet{Miville2017} concluded that the observed velocity dispersion is more likely caused by gas accretion and collapse than stellar feedback, as most of the observed massive clouds leak massive young stars, and the clouds that contain abundant O stars do not exhibit a larger turbulent velocity. A similar conclusion can be found in \citet{Traficante2018}. In the simulation, \citet{IBANEZ2016} found that the feedback from supernovae in the diffuse interstellar medium could only drive turbulent motions in dense clouds under $1\ \rm km\ s^{-1}$, which can't explain the observation. The hierarchical collapse of clouds, which converts the gravitational potential energy into kinetic energy, could primarily cause the observed fast motion within molecular clouds and maintain the balance between gravitational and kinetic energies \citep{Ballesteros-Paredes2011, Ibanez2017}.

Simulations proposed that the stellar feedback could inject energy into molecular clouds at a small scale and flatten the velocity dispersion-size relation \citep{Grisdale2018, Colman2024}, which is inconsistent with the steep slope regime. Thus, the stellar feedback may be one of the energy sources of the observed velocity dispersion in molecular clouds, but not the mechanism that leads to the steep velocity dispersion-size relation at cloud scale larger than several parsecs.

We propose that the steep slope of the velocity dispersion-size relation is not caused by some single mechanism but is determined by the interplay of shear and gravity. According to the contribution ratio between shear and self-gravity, the slope of the velocity dispersion-size relation can range from 0.5 to 1, which is determined by the intrinsic density structure of clouds and the shear rate of galactic rotation. Gravity-dominated clouds should exhibit a relatively flatter velocity dispersion-size relation with a slope close to 0.5, while shear-dominated clouds have a steeper slope.

\section{Model and application}

\subsection{Two-component model: self-gravity and shear}
\label{model}

Considering the contribution from self-gravity and shear, the velocity dispersion of molecular clouds can be estimated as follows:
\begin{equation}
\sigma_{\rm v_{total}} = \lambda_1 \sigma_{\rm v_{g}} + \lambda_2 \sigma_{\rm v_{shear}} = A[(\frac{GM}{R})^{\frac{1}{2}} + f(\frac{R}{t_{\rm shear}})]
\label{eqn:model}
\end{equation}
where $A=\lambda_1$ and $f=\lambda_2/ \lambda_1$ are two fitting parameters ($A$ determines the normalization and $f$ determines the slope of the relation), $\sigma_{\rm v_{g}}$ and $\sigma_{\rm v_{shear}}$ are the velocity dispersion contributed by self-gravity and shear, $M$ and $R$ are the mass and size of clouds, $t_{\rm shear}$ is the shear timescale, $G$ is the gravitational constant. We do not expect to derive the same values of parameters $A$ and $f$ across different observational results, as they are influenced by effects such as cloud boundary definition, cloud size definition, cloud identification, and observation tracer. We aim to obtain the specific configuration of $A$ and $f$ case-by-case. For each observation sample, we fit the observed velocity dispersion using Equation (\ref{eqn:model}) with cloud mass, size, and shear timescale, where $A$ and $f$ serve as two independent variables.

The details of $\sigma_{\rm v_{g}}$ and $\sigma_{\rm v_{shear}}$ are listed below. The gravitational potential energy of molecular clouds is $U_{\rm g} \sim {GM^2}/{R}$, and the kinetic energy converted from gravitational energy is proportional to $U_{g}$ \citep{Ballesteros-Paredes2011, IBANEZ2016}. The velocity dispersion caused by cloud self-gravity is defined as
\begin{equation}
\sigma_{\rm v_{g}} = f_1(\frac{GM}{R})^{\frac{1}{2}}
\label{eqn:v_g}
\end{equation}
where $f_1$ is the conversion factor contained within parameter $A$. 

We define the velocity dispersion caused by galactic shear as
\begin{equation}
\sigma_{\rm v_{shear}} = f_2(\frac{R}{t_{\rm shear}})
\label{eqn:v_shear}
\end{equation}
where $f_2$ is the conversion factor contained within parameters $A$ and $f$, $t_{\rm shear}$ is the shear timescale determined by
\begin{equation}
t_{\rm shear} = \kappa^{-1} = (r \left | \frac{d\Omega}{dr} \right |)^{-1} = (2A_{\rm Oort})^{-1}
\label{eqn:t_shear}
\end{equation}
where $\kappa$ is the shear rate and $A_{\rm Oort}$ is the Oort constant. In the solar vicinity, we have $t_{\rm shear} \approx 32.4 \ \rm Myr$, where we adopt $A_{\rm Oort} = 15.1 \pm 0.1\rm \ km \ s^{-1} \ {kpc}^{-1}$ derived by \citet{Li2019oort}.

\begin{figure*}
\centering
\includegraphics[width=\textwidth]{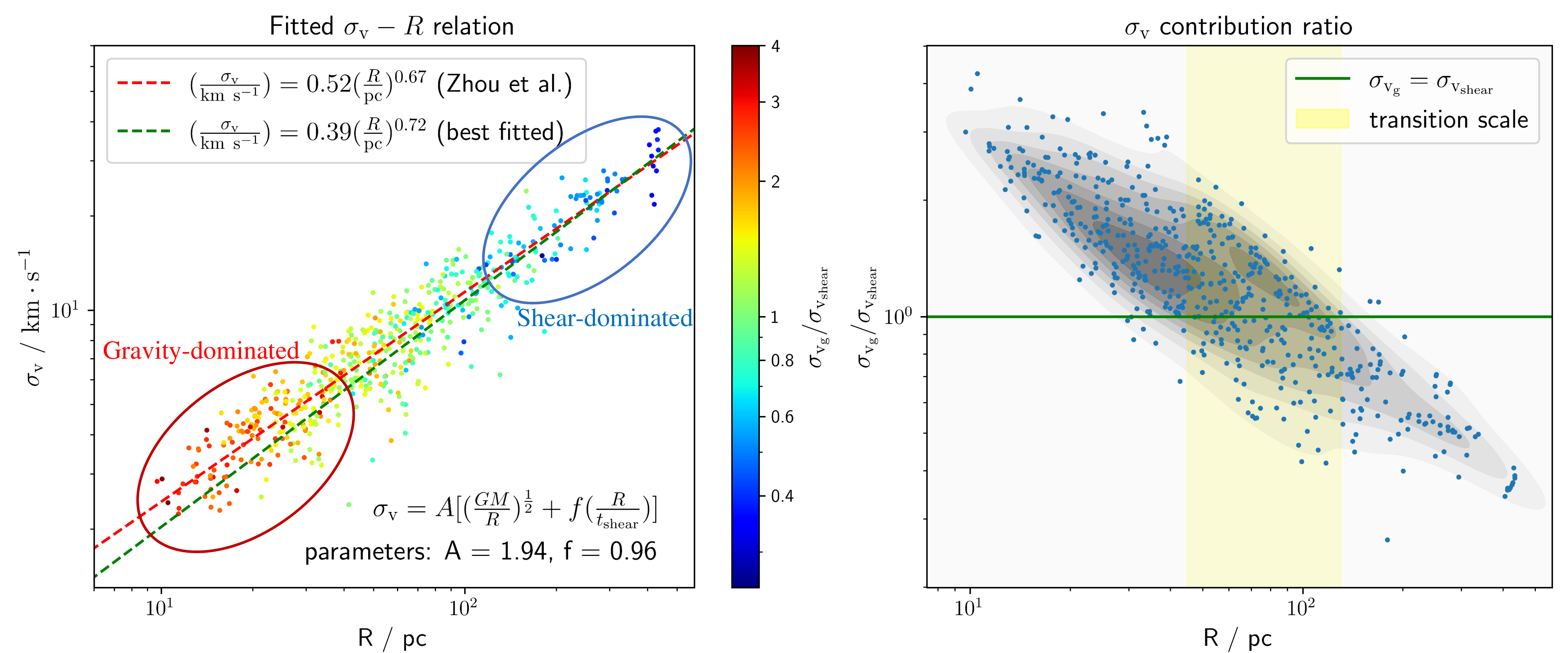}
\caption{\textbf{Left: Fitted velocity dispersion-size relation in the solar vicinity.} The red dashed line shows the one-dimensional $\sigma_{\rm v}-R$ relation derived from \citet{zhou2022}. The scatters show the corresponding velocity dispersion of clouds from \citet{Xie2024} calculated via Equation (\ref{eqn:v_zhou_1D}) with $20\%$ Gaussian noise. The green dashed line shows the estimating result with the parameters $A=1.94$ and $f=0.96$ of Equation (\ref{eqn:model}). The color of scatters represents the $\sigma_{\rm v}$ contribution ratio between self-gravity and shear calculated via Equation (\ref{eqn:ratio}). The red and blue ellipses show the gravity-dominated and shear-dominated regions, respectively. \textbf{Right: Contribution ratio between self-gravity and shear as a function of cloud size.} The green line shows where self-gravity and shear contribute equally to velocity dispersion. The yellow region represents the transition scale of $\sim 100\ \rm pc$. The background gray colormap shows the kernel density estimation result.}
\label{fig:Zhou}
\end{figure*}

\begin{figure*}
\centering
\includegraphics[width = 6 in]{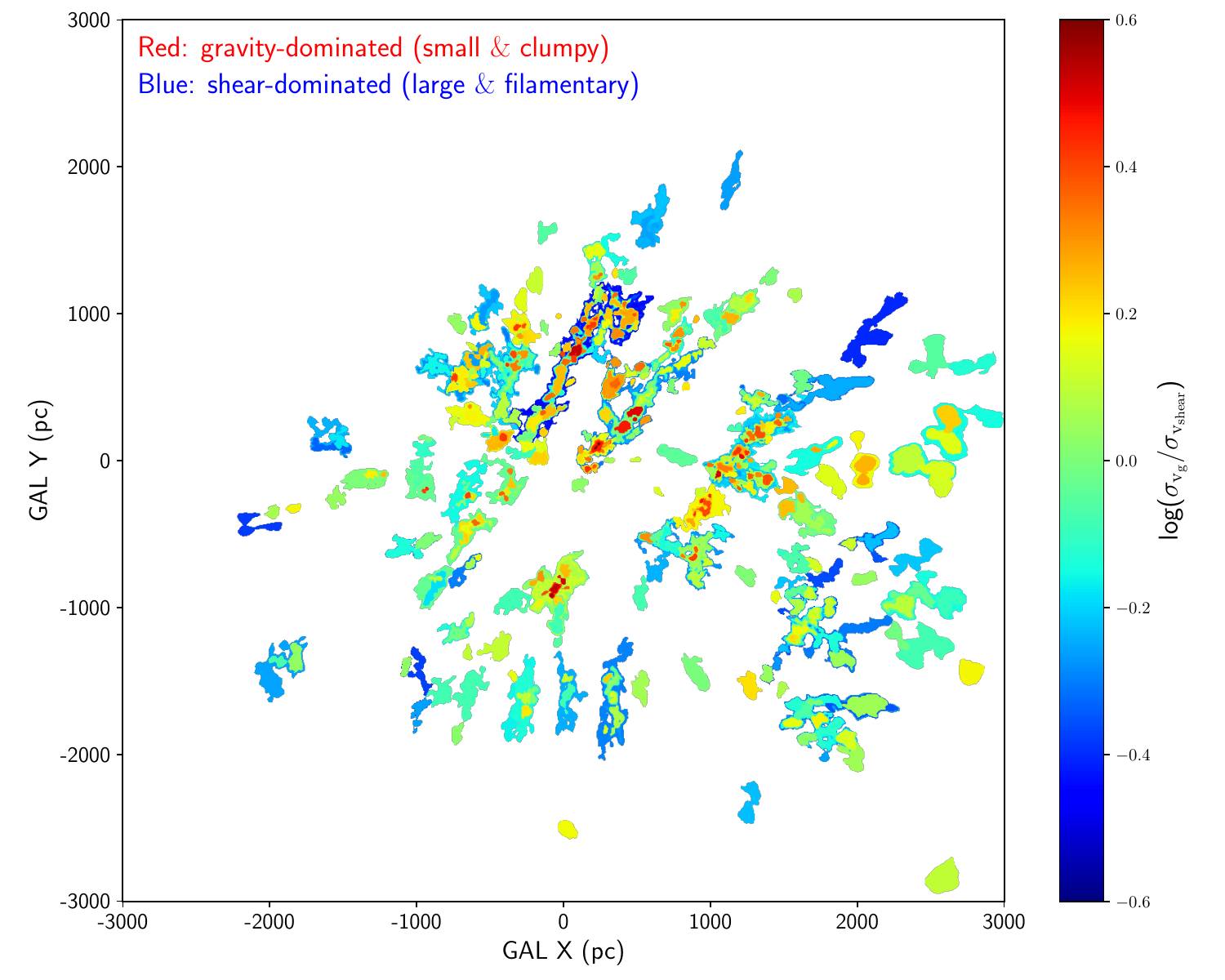}
\caption{\textbf{Locations of shear-dominated and gravity-dominated clouds in the Milky Way disk.} The map shows the cloud spatial distribution projected on the Galactic mid-plane (X-Y plane). The color of clouds refers to the velocity dispersion contribution ratio between gravity and shear calculated via Equation (\ref{eqn:ratio}).}
\label{fig:mask}
\end{figure*}

\subsection{Steep velocity dispersion-size relation in the solar vicinity}
\label{Zhou}

\subsubsection{Observation data}
\label{data}

We adopt the observation sample from \citet{zhou2022} and \citet{Xie2024} to determine the full form of our model. \citet{zhou2022} calculated the velocity dispersion and size of their identified 150 YSO associations within $d_{\rm sun} \lesssim 3$ kpc based on astrometric measurement data from \textit{Gaia} DR2. They derived a steep two-dimensional velocity dispersion-size relation of $\sigma_{\rm v_{(zhou)}}(\rm 2D) = 0.74\ (R/\rm pc)^{0.67}\ (km\ s^{-1})$. Since \citet{zhou2022} did not measure the mass of their YSO associations, we further use the cloud sample from \citet{Xie2024}, which identified 550 cloud structures in the solar vicinity from the 3D dust extinction map derived by \citet{Vergely2022} with their mass and size. The YSO association sample and the cloud sample from these two works are located within a similar region in the solar vicinity and exhibit a similar size range from a few to hundreds of parsec, thus we propose that the velocity dispersion-size relation derived by \citet{zhou2022} fits with the cloud sample from \citet{Xie2024}.

\textbf{Size calibration:} We calibrate the cloud size from \citet{Xie2024} to match with the size definition of \citet{zhou2022}. \citet{Xie2024} calculated the cloud size ($R_{\rm max}$, $R_{\rm mid}$, $R_{\rm min}$) based on three eigenvalues of the 3D inertia tensor of each cloud, while \citet{zhou2022} derived the size of YSO associations from the 2D spatial distribution dispersion on the plane-of-sky ($r=2.355d\cdot \sqrt{(\sigma_{\rm max}^2 + \sigma_{\rm min}^2)/2} $). We calculate the projected size on the plane-of-sky of sample clouds from \citet{Xie2024} as follows:
\begin{equation}
R_{\rm project}=\sqrt{[(\cos{45^\circ})\cdot R_{\rm max}]^2 + (R_{\rm mid})^2}
\label{eqn:size calibration}
\end{equation}
where we assume a typical angle of $45^\circ$ between the cloud major axes and the line-of-sight.

\textbf{Velocity dispersion calibration:} To compare the two-dimensional velocity dispersion from \citet{zhou2022} with other CO-traced one-dimensional velocity dispersion along the line-of-sight, we calculate their corresponding one-dimensional velocity dispersion by assuming $\sigma_{\rm v_{2D}} = \sqrt{2}\ \sigma_{\rm v_{1D}}$ as follows:

\begin{equation}
\sigma_{\rm v_{(zhou)}}(\rm 1D) = 0.52\ (R/\rm pc)^{0.67}\ (km\ s^{-1})
\label{eqn:v_zhou_1D}
\end{equation}

\subsubsection{Model fitting}
\label{Zhou_fit}

We apply our model to the observation result from \citet{zhou2022} and \citet{Xie2024} to determine its full form and demonstrate its reliability. To link the velocity dispersion-size relation from \citet{zhou2022} with the cloud mass and size from \citet{Xie2024}, we apply Equation (\ref{eqn:v_zhou_1D}) to the calibrated cloud size from Equation (\ref{eqn:size calibration}) to calculate the corresponding velocity dispersion (we add a Gaussian noise of 20\% to represent observational uncertainties). We use Equation (\ref{eqn:model}) to fit this corresponding velocity dispersion with the cloud mass and calibrated size from \citet{Xie2024} and the shear timescale in section \ref{model}. The best-fitted parameters are $A=1.94$ and $f=0.96$. We substitute the fitted parameters into Equation (\ref{eqn:model}) to calculate the estimated velocity dispersion of each cloud. The fitting relation between the estimated velocity dispersion with cloud size is
\begin{equation}
\sigma_{\rm v_{estimate}} = 0.39\ (R/\rm pc)^{0.72}\ (km\ s^{-1})
\label{eqn:v_fit}
\end{equation}
Our estimating result is shown in the left panel of Fig. \ref{fig:Zhou}. The fitted slope of 0.72 agrees with the original one of 0.67. The fitted value of $f$ is close to 1, indicating that self-gravity and shear have a similar contribution efficiency to velocity dispersion.

\subsubsection{Gravity-shear transition}

We investigate the velocity dispersion contribution ratio between self-gravity and shear of each cloud in section \ref{Zhou_fit} as follows:
\begin{equation}
\frac{\lambda_1\sigma_{\rm v_g}}{\lambda_2\sigma_{\rm v_{shear}}} = \frac{t_{\rm shear}}{f}(\frac{GM}{R^3})^{\frac{1}{2}}
\label{eqn:ratio}
\end{equation}
where $f$ is the parameter in Equation (\ref{eqn:model}). As shown in the right panel of Fig. \ref{fig:Zhou}, the velocity dispersion of small clouds is mainly caused by self-gravity, while large clouds are primarily affected by shear. The transition scale is $\sim100\ \rm pc$---nearly the scale height of the Galactic molecular gas disk. This scale is consistent with the shape transition scale between small ellipsoid-shaped clouds and large filamentary structures in \citet{Xie2024}. We plot our result with the cloud shape projected on the Galactic mid-plane (X-Y plane) in Fig. \ref{fig:mask}. Gravity-dominated clouds are small and clumpy, while shear-dominated clouds are large and filamentary. These results indicate that the steep velocity dispersion-size relation above several parsecs reflects a gradual transition from a gravity-dominated to a shear-dominated mechanism.

\subsubsection{Virial parameters}

\begin{figure}
\centering
\includegraphics[width=\columnwidth]{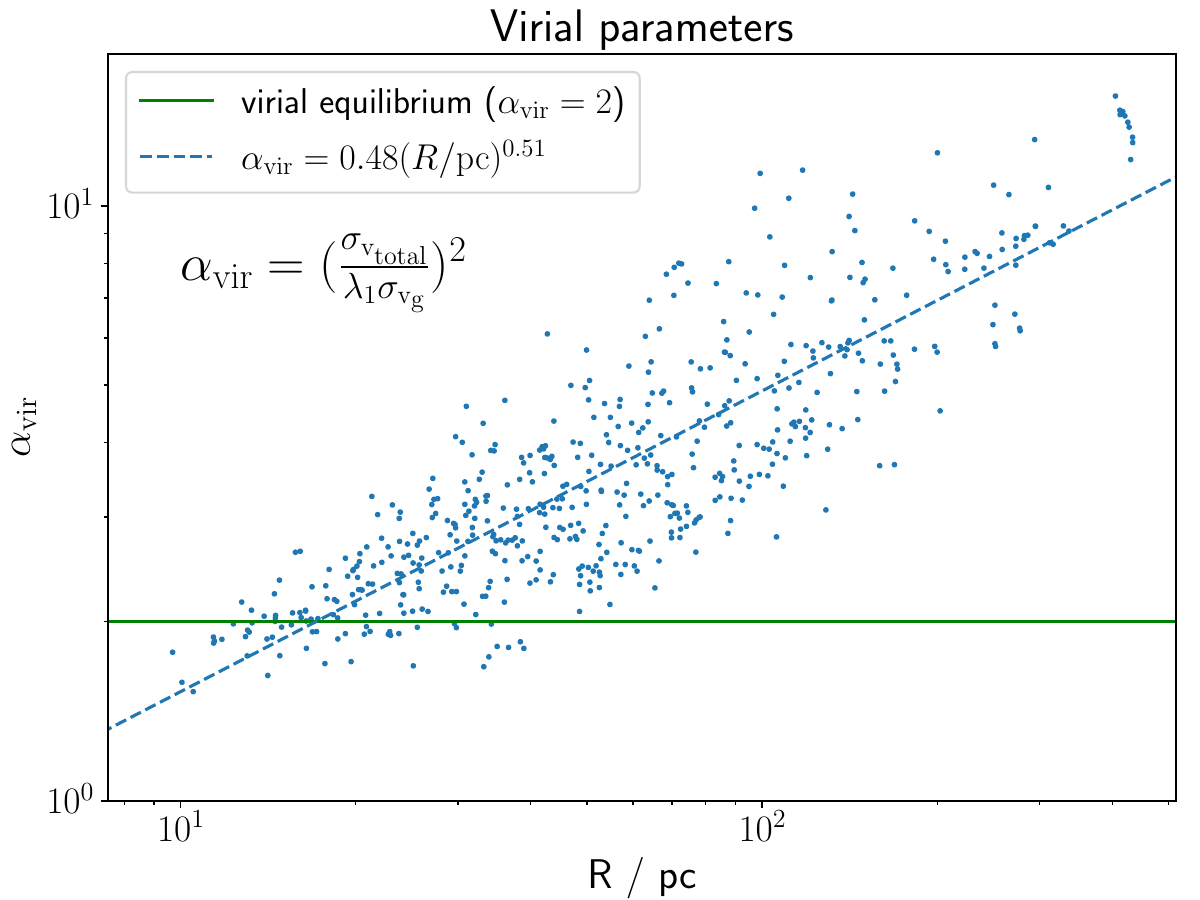}
\caption{\textbf{Virial parameter as a function of cloud size.} The blue dashed line shows the fitting result. The green solid line represents where clouds are in virial equilibrium ($\alpha_{\rm vir} = 2$).}

\label{fig:virial}
\end{figure}

We investigate the virial parameters of sample clouds in section \ref{Zhou_fit} as follows:
\begin{equation}
\alpha_{\rm vir} = (\frac{\sigma_{\rm v_{total}}}{\lambda_1 \sigma_{\rm v_g}})^2
\label{eqn:virial}
\end{equation}
where $\sigma_{\rm v_{total}}$ is the estimated total velocity dispersion and $\lambda_1 \sigma_{\rm v_g}$ is the self-gravity contributed velocity dispersion from Equation (\ref{eqn:model}). As shown in Fig. \ref{fig:virial}, the virial parameter increases with cloud size, where the fitting result is as follows:
\begin{equation}
\alpha_{\rm vir} = 0.48(R/\rm pc)^{0.51}
\label{eqn:virial-R}
\end{equation}
According to the commonly used criteria $\alpha_{\rm vir}\sim2$ for virial equilibrium\citep{Kauffmann2013}, small clouds are roughly in virial equilibrium, while large structures are supervirial. This result is consistent with the simulation of \citet{Dobbs2011}, where the most part of the molecular clouds are not gravitationally bound.

Our result is contradictory to previous studies where the virial parameter declines with cloud size (e.g. \citealt{Traficante2018, Benedettini2020}) or cloud mass (e.g. \citealt{Kauffmann2013, Miville2017, Chevance2023, Sun2024}). Our explanation is that the traditional virial parameter-size or virial parameter-mass relation embodies two relations, namely the mass-size relation and velocity dispersion-size relation, both of which depend on the definition of cloud boundary and size. In our work, following the calibrated cloud size from Equation (\ref{eqn:size calibration}), we have the mass-size relation as follows:
\begin{equation}
(M/M_{\odot})=19.01\ (R/ \rm pc)^{1.94}
\label{eqn:mass-size}
\end{equation}
According to the definition of virial parameter $\alpha_{\rm vir}\propto \sigma_{\rm v}^2 R/M$ \citep{McKee1992}, following the Appendix B of \citet{Kauffmann2013}, the relation between the slope of $M\propto R^{\beta_1}$, $\sigma_{\rm v}\propto R^{\beta_2}$ and $\alpha_{\rm vir}\propto R^{\beta_3}$ relations is
\begin{equation}
\beta_3 = 2\beta_2 - \beta_1 + 1
\label{eqn:virial_slope}
\end{equation}
The positive slope of virial parameter-size relation should hold for our sample clouds with size larger than 10 pc, since our $M\propto R^{1.94}$ and $\sigma_{\rm v}\propto R^{0.72}$ (Equation (\ref{eqn:v_fit})) relations necessarily imply such positive virial parameter-size relation with a slope $\sim 0.5$.

\subsection{Prediction of velocity dispersion-size relation at different Galactocentric distances}
\label{Miville}

\begin{figure}
\centering
\includegraphics[width=\columnwidth]{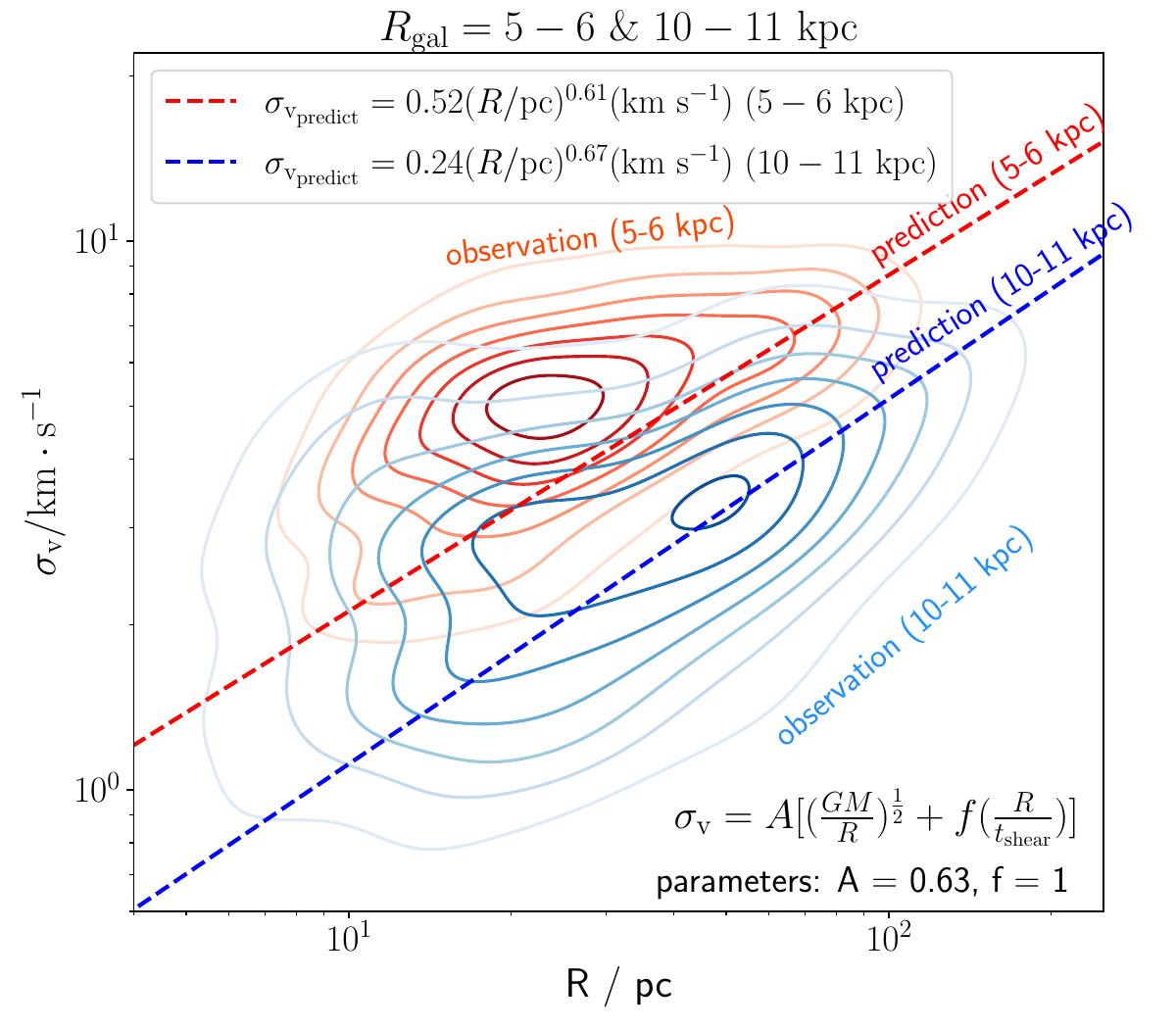}
\caption{\textbf{Prediction of the variation of velocity dispersion at different Galactocentric distances.} The red ($R_{\rm gal}=5-6$ kpc) and blue ($R_{\rm gal}=10-11$ kpc) contours show the kernel density estimation of the observation result from \citet{Miville2017}. The red ($R_{\rm gal}=5-6$ kpc) and blue ($R_{\rm gal}=10-11$ kpc) dashed lines represent our prediction result using Equation (\ref{eqn:model}).}
\label{fig:Miville}
\end{figure}

As an application and reliability verification of our model on explaining the variation of velocity dispersion-size relation of molecular clouds within the Galactic disk, we adopt the observation result containing cloud size, mass, and velocity dispersion measurements from \citet{Miville2017}, which identified 8107 molecular clouds covering the entire Galactic disk plane based on the CO survey data from \citet{Dame2001}. Their cloud sample exhibits a size range from a few tenths to hundreds of parsecs.

Following the description in section \ref{model}, we do not expect to obtain a similar value of parameter $A$ of Equation (\ref{eqn:model}) with the result in section \ref{Zhou_fit}. However, we expect that the value of parameter $f$ should be similar here since the slope of the velocity dispersion-size relation obtained by \citet{zhou2022} of 0.67 and \citet{Miville2017} of $\sim 0.63$ are similar.

\citet{Miville2017} derived a velocity dispersion-size relation of their whole cloud sample as follows: 
\begin{equation}
\sigma_{\rm v_{(Miville)}} = 0.48\ (R/\rm pc)^{0.63\pm 0.30}\ (km\ s^{-1})
\label{eqn:v_miville}
\end{equation}
which is similar to the one derived from \citet{zhou2022} (Equation \ref{eqn:v_zhou_1D}). We assume $f=1$ following the result $f = 0.96$ in section \ref{Zhou_fit} and apply Equation (\ref{eqn:model}) to the cloud sample of \citet{Miville2017} with Galactocentric distance $4<R_{\rm gal}<20$ kpc to derive an appropriate value of $A$. We exclude clouds with $R_{\rm gal}<4$ kpc and $R_{\rm gal}>20$ kpc as their rotation curves are not accurately measured. We calculate the shear timescale at different Galactocentric distances via Equation (\ref{eqn:t_shear}) using the linear Galactic rotation curve model obtained by \citet{Mroz2019}. \citet{Miville2017} calculated the cloud mass via a constant factor $X_{\rm CO(const)}=2\times 10^{20}\rm \ cm^{-2}\ K^{-1}\ km^{-1}\ s$ \citep{Bolatto2013}. We calibrate the cloud mass according to the $X_{\rm CO}$ factor calibration from \citet{Lada2020} as follows:
\begin{equation}
X_{\rm CO}(R_{\rm gal})=
\begin{cases}
   (\frac{83}{54.5-3.7R_{\rm gal}})\ (2<R_{\rm gal}\le10)\\
   6.0\hspace{4.8em} (R_{\rm gal}>10)
\end{cases}
\label{eqn:XCO}
\end{equation}
where $X_{\rm CO}$ is in units of $10^{20}\ \rm cm^{-2}\ K^{-1}\ km^{-1}\ s$ and $R_{\rm gal}$ is in units of kpc. The cloud mass is calibrated by:
\begin{equation}
M_{\rm calibrate} = M_{(\rm Miville)} \frac{X_{\rm CO(calibrate)}}{X_{\rm CO(const)}}
\label{eqn:re-mass}
\end{equation}

Through the similar procedure in section \ref{Zhou_fit}, we fit the observed velocity dispersion using Equation (\ref{eqn:model}). Our fitting result is $A=0.63$ corresponding to 
\begin{equation}
\sigma_{\rm v_{estimate}} = 0.38\ R^{0.61}\ (4<R_{\rm gal}<20\ \rm kpc)
\label{eqn:v_4-16}
\end{equation}
where $\sigma_{\rm v}$ is in units of $\rm km\ s^{-1}$ and $R$ is in units of pc. Our estimated velocity dispersion-size relation has a similar slope but a slightly smaller normalization than the original result in Equation (\ref{eqn:v_miville}). This deviation may be because we exclude clouds with Galactocentric distance $R_{\rm gal}<4$ kpc, which exhibits a significantly larger average velocity dispersion.

We use Equation (\ref{eqn:model}) with derived parameters ($A=0.63,\ f=1$) and cloud size and mass to estimate the velocity dispersion-size relation at different Galactocentric distances. We extract two groups of sample clouds from \citet{Miville2017} located within (1) $5<R_{\rm gal}<6$ kpc and (2) $10<R_{\rm gal}<11$ kpc. These clouds have similar solar distances, which can reduce the observation effect that the fraction of observed small clouds declines with solar distance. Fig. \ref{fig:Miville} shows the fitted relation of estimated velocity dispersion. For these clouds we have 
\begin{equation}
\sigma_{\rm v_{predict}}=
\begin{cases}
   0.52\ R^{0.61}\ (5<R_{\rm gal}<6\ \rm kpc)\\
   0.24\ R^{0.67}\ (10<R_{\rm gal}<11\ \rm kpc)
\end{cases}
\label{eqn:5-6&10-11}
\end{equation}
where $\sigma_{\rm v}$ is in units of $\rm km\ s^{-1}$ and $R$ is in units of pc. The observation data exhibit a large scatter at a small scale in the $\sigma_{\rm v}-R$ space, making it difficult to constrain the slope of the velocity dispersion-size relation. However, our model agrees well with observation on a large scale. (Since \citet{Miville2017} did not calculate the slope of velocity dispersion-size relation at different Galactocentric distances, we adopt the observation result from \citet{Sun2024} to demonstrate the reliability of our model on predicting the slope at different Galactocentric distances in Appendix \ref{Appendix A}.)

\citet{Miville2017} calculated the average velocity dispersion of clouds is $\sim 5\ \rm km\ s^{-1}$ at $5<R_{\rm gal}<6$ kpc and $\sim 3\ \rm km\ s^{-1}$ at $10<R_{\rm gal}<11$ kpc. Our predicted average velocity dispersion at the same distances is $4.2\ \rm km\ s^{-1}$ and $2.7\ \rm km\ s^{-1}$ respectively, which is consistent with the observation. In our model, the variation of predicted velocity dispersion-size relation results from the variation of cloud density structure (which determines the cloud self-gravity) and shear timescale. Thus we propose that the variation of these two factors determines the variation of cloud velocity dispersion-size relation with the Galactocentric distance, since both the cloud surface density \citep{Miville2017} and shear rate decline with Galactocentric distance from $R_{\rm gal}=5$ kpc to $R_{\rm gal}=11$ kpc.

From Fig. \ref{fig:Miville}, we find that the observed cloud velocity dispersion at a small scale is larger than our predicted power-law relation. This is consistent with the observation result that the velocity dispersion-size relation becomes flatter or even decoupled from cloud size under several parsecs ($\sim2$ pc for \citet{Benedettini2020, Sun2024}, $\sim7$ pc for \citet{Heyer2001, Miville2017}, also in \citet{Traficante2018, Luo2024}). This break may be caused by the energy injection from stellar feedback at a small scale \citep{Colman2024}. The mechanisms that control the cloud velocity dispersion under several parsecs need further investigation.

\section{Conclusion}

\begin{figure*}
\centering
\includegraphics[width=\textwidth]{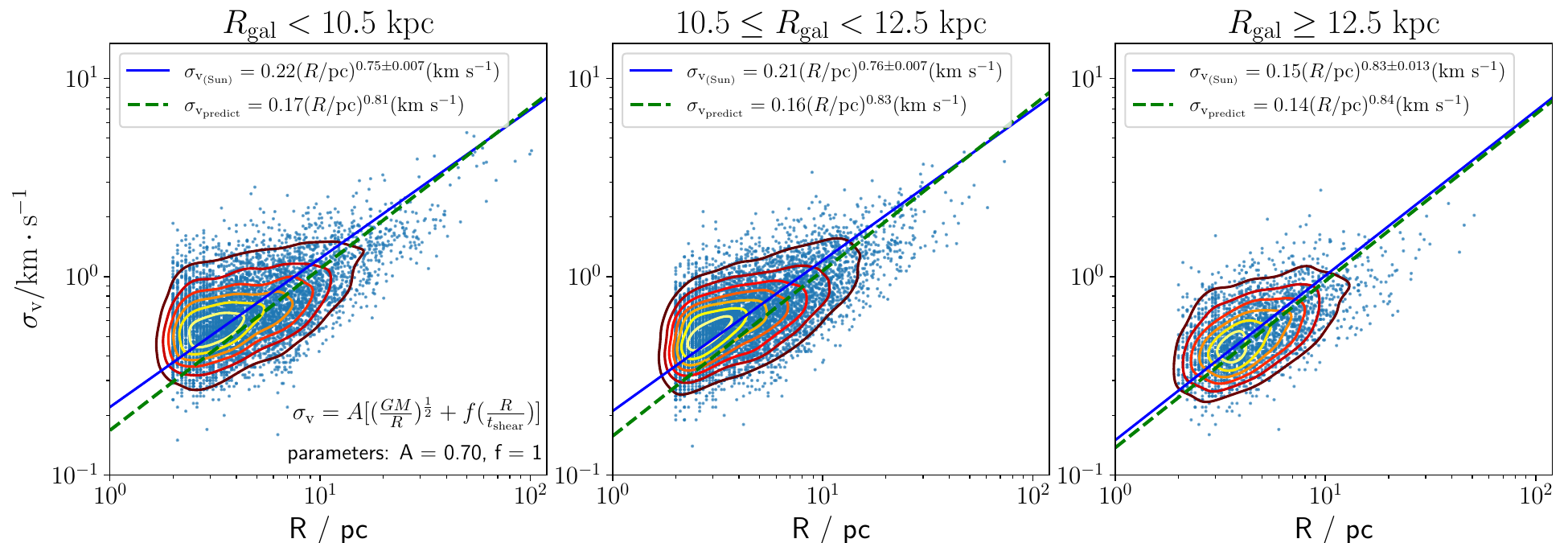}
\caption{\textbf{Prediction of the slope of velocity dispersion-size relation at different Galactocentric distances.} The left, middle, and right panels show the observation and prediction result at $R_{\rm gal}<10.5$ kpc, $10.5\leq R_{\rm gal}<12.5$ kpc, and $R_{\rm gal}\geq 12.5$ kpc, respectively. The blue scatters and the contours show the observation result from \citet{Sun2024}, and the blue solid lines represent their fitting results. The green dashed lines represent our prediction results.}
\label{fig:Sun}
\end{figure*}

The velocity dispersion-size relation is a fundamental relation describing the property of interstellar gas, whose slope reflects the physical mechanisms controlling its evolution. In previous studies, this slope is considered as $\sim 0.5$, which can be explained by virial equilibrium or compressible supersonic turbulence. Recently, studies found a steeper slope of $\beta \sim 0.6-0.8$, which a single mechanism with only gravity or shear cannot explain. We present a two-component model including shear and self-gravity as follows:
\begin{equation}
\sigma_{\rm v_{total}} = \lambda_1 \sigma_{\rm v_{g}} + \lambda_2 \sigma_{\rm v_{shear}} = A[(\frac{GM}{R})^{\frac{1}{2}} + f(\frac{R}{t_{\rm shear}})]
\end{equation}
to explain the steep velocity dispersion-size relation in observation for molecular clouds larger than several parsecs. Our model agrees well with the observation result in the solar vicinity derived by \citet{zhou2022} (Fig. \ref{fig:Zhou}). We further predict the variation and slope of cloud velocity dispersion-size relation with different Galactocentric distances in the observation of \citet{Miville2017} (Fig. \ref{fig:Miville}) and \citet{Sun2024} (Fig. \ref{fig:Sun}), respectively. 

Our results indicate that the transition from the gravity-dominated to the shear-dominated regime offers a simple explanation for the steep velocity dispersion-size relation observed toward clouds above several parsecs in different environments. The velocity dispersion of small clouds is mainly caused by self-gravity, while large clouds are primarily influenced by shear, where the transition scale is $\sim100$ pc in the solar vicinity---the scale height of the Galactic molecular gas disk. The virial parameter of our cloud sample increases with cloud size. Small clouds are roughly in virial equilibrium, while large structures are supervirial. We propose that the variation of cloud velocity dispersion-size relation with the Galactocentric distance results from the variation of cloud density structure and shear rate, which determine cloud self-gravity and galactic shear, respectively. Our model enables us to investigate how conditions, including cloud density structure and galactic shear, influence the velocity dispersion of molecular clouds in future surveys and simulations.

\begin{acknowledgments}

Guang-Xing Li acknowledges support from NSFC grant Nos. 12273032 and 12033005.

\end{acknowledgments}

\vspace{5mm}
\facilities{Gaia, FLWO:2MASS, CTIO:2MASS, MWISP, MWT}

\software{astropy \citep{astropy:2013, astropy:2018, astropy:2022}
          BCES \citep{Akritas1996, Nemmen2012}
          Dendrogram \citep{Rosolowsky2008}
          glue \citep{glue2015, glue2017}
          numpy \citep{numpy2020}
          scipy \citep{scipy2020}
          }

\appendix

\begin{figure*}
\centering
\includegraphics[width = \textwidth]{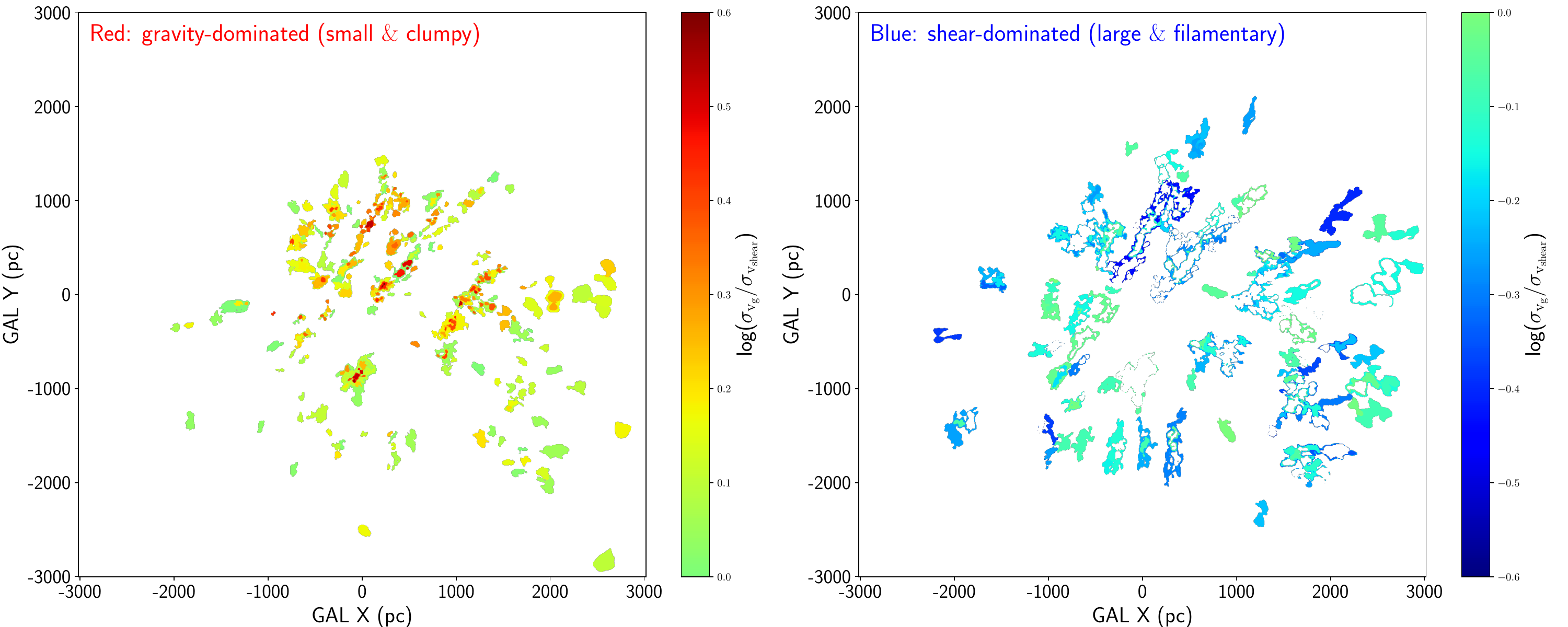}
\caption{\textbf{Map of gravity-dominated and shear-dominated gas.} Left panel: gravity-dominated gas ($\lambda_1 \sigma_{\rm v_g} > \lambda_2 \sigma_{\rm v_{shear}}$); right panel: shear-dominated gas ($\lambda_1 \sigma_{\rm v_g} < \lambda_2 \sigma_{\rm v_{shear}}$). The clouds are projected on the mid-plane of the Galactic disk. The color of clouds refers to the velocity dispersion contribution ratio between gravity and shear from Equation (\ref{eqn:ratio}).}
\label{fig:mask_g_s}
\end{figure*}

\begin{figure*}
\centering
\includegraphics[width = 6in]{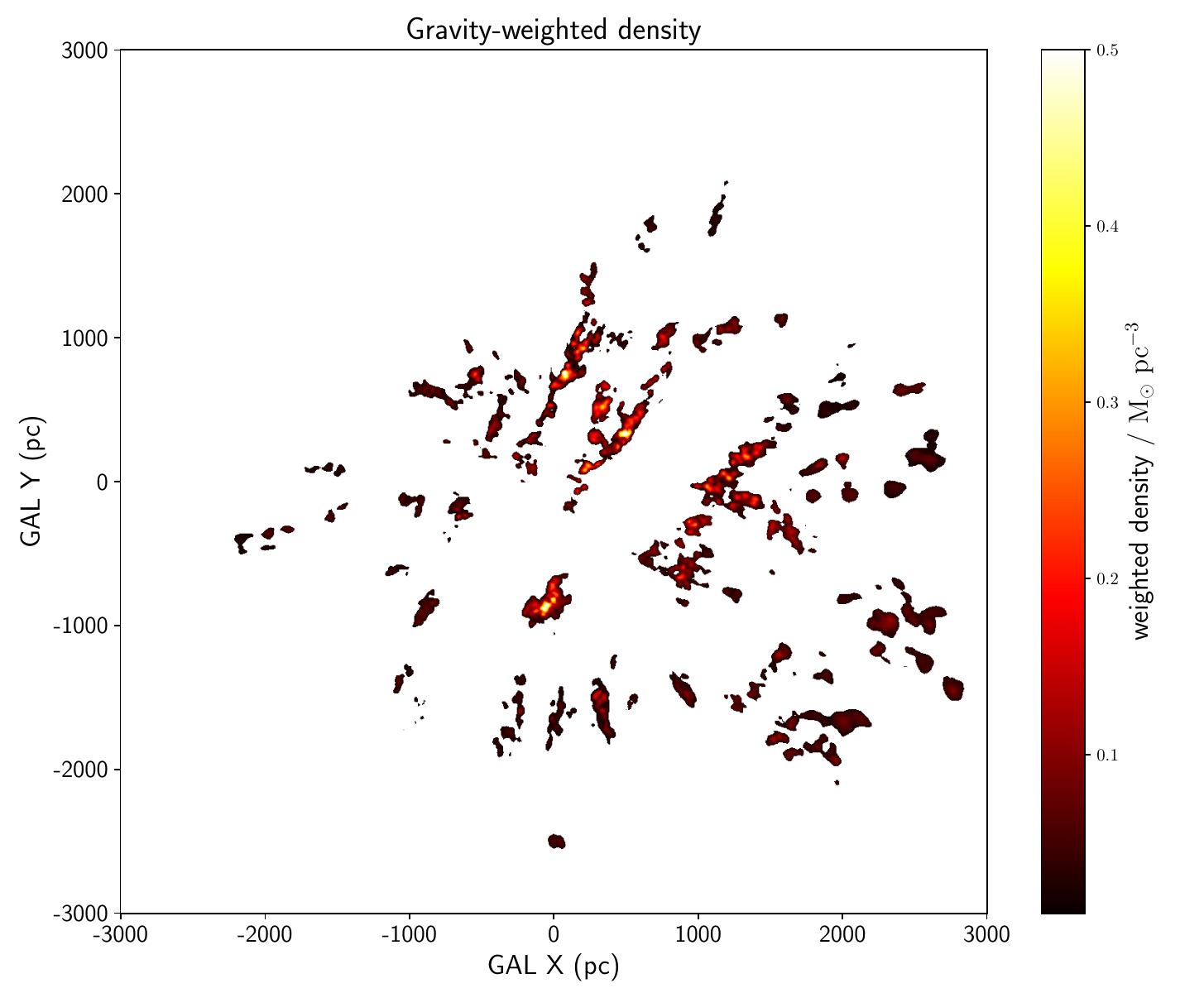}
\caption{\textbf{Gravity-weighted gas density at Galactic mid-plane ($z=0$).}}
\label{fig:rho_g}
\end{figure*}

\begin{figure*}
\centering
\includegraphics[width = 6in]{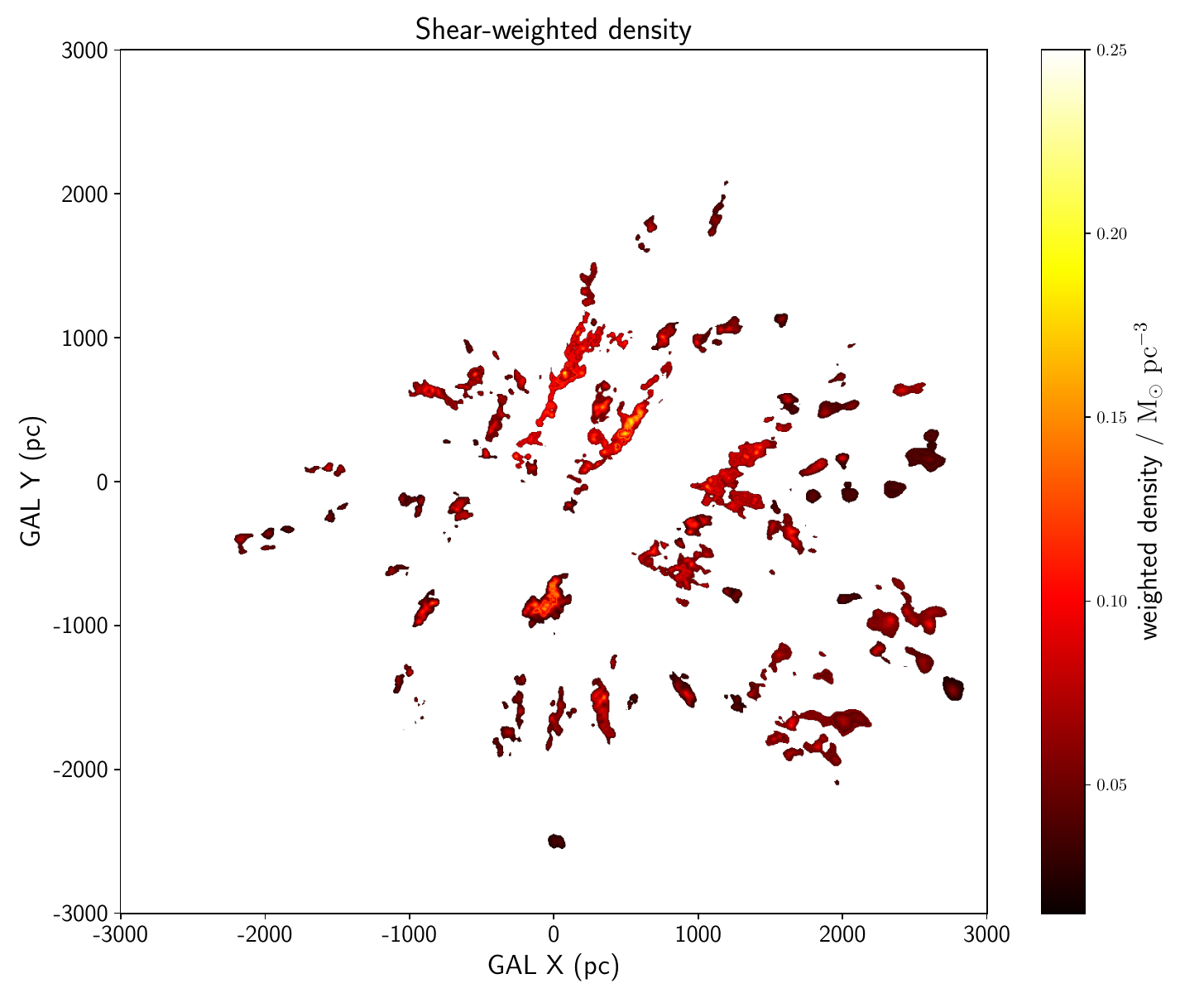}
\caption{\textbf{Shear-weighted gas density at Galactic mid-plane ($z=0$).}}
\label{fig:rho_s}
\end{figure*}

\section{Prediction of the slope of velocity dispersion at different Galactocentric distances}
\label{Appendix A}

We use the cloud sample from \citet{Sun2024} to demonstrate the reliability of our shear-gravity model in predicting the slope of velocity dispersion at different Galactocentric distances. \citet{Sun2024} identified 32162 molecular clouds within the northern outer Galactic disk with Galactocentric distance across $8<R_{\rm gal}<26$ kpc based on the $^{12}\rm CO$ J=1-0 emission line from MWISP project. Their cloud sample has a size range from a few tenths to a hundred parsecs.

\citet{Sun2024} obtained a velocity dispersion-size relation of the resolved cloud sample with cloud size $\geq 2$ pc as follows:
\begin{equation}
\sigma_{\rm v_{(Sun)}} = 0.19\ (R/\rm pc)^{0.78\pm 0.005}\ (km\ s^{-1})
\label{eqn:sun_total}
\end{equation}
where the slope is larger than the result with the whole resolved sample $\beta=0.65\pm 0.004$. They also calculated the relations at different Galactocentric distances with size $\geq 2$ pc as follows:
\begin{equation}
\sigma_{\rm v_{(Sun)}}=
\begin{cases}
   0.22\ R^{0.75\pm 0.007}\ (R_{\rm gal}<10.5\ \rm kpc)\\
   0.21\ R^{0.76\pm 0.007}\ (10.5\leq R_{\rm gal}<12.5\ \rm kpc)\\
   0.15\ R^{0.83\pm 0.013}\ (R_{\rm gal}\geq 12.5\ \rm kpc)
\end{cases}
\label{eqn:Sun_bin}
\end{equation}
where $\sigma_{\rm v}$ is in units of $\rm km\ s^{-1}$ and $R$ is in units of pc.

With the same procedure in section \ref{Miville}, we assume $f=1$ and extract the resolved clouds from \citet{Sun2024} with $R_{\rm gal} < 20$ kpc and cloud size $> 2$ pc to fit the appropriate value of $A$. We exclude clouds smaller than 2 pc as \citet{Sun2024} reported that the velocity dispersion-size relation becomes flattened with clouds smaller than $\sim 2$ pc. To match their fitting results, we use the BCES bisector method \citep{Akritas1996, Nemmen2012} following \citet{Sun2024} to calculate the velocity dispersion-size relation in this section. Our fitting result is $A = 0.70$, corresponding to
\begin{equation}
\sigma_{\rm v_{estimate}} = 0.16\ (R/\rm pc)^{0.83}\ (km\ s^{-1})
\label{eqn:sun_fit}
\end{equation}
Our estimated relation agrees with Equation (\ref{eqn:sun_total}) well. With the derived parameters $A=0.70$ and $f=1$, we use Equation (\ref{eqn:model}) with cloud size, mass, and shear timescale to estimate the velocity dispersion at different Galactocentric distances. Our prediction results are shown in Fig. \ref{fig:Sun} and as follows:
\begin{equation}
\sigma_{\rm v_{predict}}=
\begin{cases}
   0.17\ R^{0.81}\ (R_{\rm gal}<10.5\ \rm kpc)\\
   0.16\ R^{0.83}\ (10.5\leq R_{\rm gal}<12.5\ \rm kpc)\\
   0.14\ R^{0.84}\ (R_{\rm gal}\geq 12.5\ \rm kpc)
\end{cases}
\label{eqn:Sun_fit_bin}
\end{equation}
where $\sigma_{\rm v}$ is in units of $\rm km\ s^{-1}$ and $R$ is in units of pc. The deviation between the predicted slopes with observed ones at different Galactocentric distances is within 0.07. This result demonstrates that our model can explain the observed steep slope of velocity dispersion-size relation for clouds larger than several parsecs within the Galaxy.

\section{Map of gravity-dominated and shear-dominated gas in the solar vicinity}
\label{Appendix B}

Using the same data from Fig. \ref{fig:mask}, we plot the spatial distribution of gravity-dominated gas ($\lambda_1 \sigma_{\rm v_g} > \lambda_2 \sigma_{\rm v_{shear}}$) and shear-dominated gas ($\lambda_1 \sigma_{\rm v_g} < \lambda_2 \sigma_{\rm v_{shear}}$) in Fig. \ref{fig:mask_g_s} respectively.

\section{Gravity-weighted and shear-weighted gas density in the solar vicinity}
\label{Appendix C}

We calculate the gravity-weighted gas density and shear-weighted gas density of the sample clouds in section \ref{Zhou} as follows:
\begin{equation}
\rho_{\rm g} = \rho_0 \frac{\lambda_1 \sigma_{\rm v_g}}{\sigma_{\rm v_{total}}}
\label{eqn:rho_g}
\end{equation}

\begin{equation}
\rho_{\rm shear} = \rho_0 \frac{\lambda_2 \sigma_{\rm v_{shear}}}{\sigma_{\rm v_{total}}}
\label{eqn:rho_s}
\end{equation}
where $\rho_0$ is the original gas density, $\sigma_{\rm total}$ is the estimated total velocity dispersion from Equation (\ref{eqn:model}), $\lambda_1 \sigma_{\rm v_g}$ is the gravity-caused velocity dispersion and $\lambda_2 \sigma_{\rm v_{shear}}$ is the shear-caused velocity dispersion. Fig. \ref{fig:rho_g} and \ref{fig:rho_s} show our result at the Galactic mid-plane ($z=0$), respectively.

\bibliography{sample631}{}

\begin{thebibliography}{}
\expandafter\ifx\csname natexlab\endcsname\relax\def\natexlab#1{#1}\fi
\providecommand{\url}[1]{\href{#1}{#1}}
\providecommand{\dodoi}[1]{doi:~\href{http://doi.org/#1}{\nolinkurl{#1}}}
\providecommand{\doeprint}[1]{\href{http://ascl.net/#1}{\nolinkurl{http://ascl.net/#1}}}
\providecommand{\doarXiv}[1]{\href{https://arxiv.org/abs/#1}{\nolinkurl{https://arxiv.org/abs/#1}}}

\bibitem[{{Akritas} \& {Bershady}(1996)}]{Akritas1996}
{Akritas}, M.~G., \& {Bershady}, M.~A. 1996, \apj, 470, 706, \dodoi{10.1086/177901}

\bibitem[{{Astropy Collaboration} {et~al.}(2013){Astropy Collaboration}, {Robitaille}, {Tollerud}, {Greenfield}, {Droettboom}, {Bray}, {Aldcroft}, {Davis}, {Ginsburg}, {Price-Whelan}, {Kerzendorf}, {Conley}, {Crighton}, {Barbary}, {Muna}, {Ferguson}, {Grollier}, {Parikh}, {Nair}, {Unther}, {Deil}, {Woillez}, {Conseil}, {Kramer}, {Turner}, {Singer}, {Fox}, {Weaver}, {Zabalza}, {Edwards}, {Azalee Bostroem}, {Burke}, {Casey}, {Crawford}, {Dencheva}, {Ely}, {Jenness}, {Labrie}, {Lim}, {Pierfederici}, {Pontzen}, {Ptak}, {Refsdal}, {Servillat}, \& {Streicher}}]{astropy:2013}
{Astropy Collaboration}, {Robitaille}, T.~P., {Tollerud}, E.~J., {et~al.} 2013, \aap, 558, A33, \dodoi{10.1051/0004-6361/201322068}

\bibitem[{{Astropy Collaboration} {et~al.}(2018){Astropy Collaboration}, {Price-Whelan}, {Sip{\H{o}}cz}, {G{\"u}nther}, {Lim}, {Crawford}, {Conseil}, {Shupe}, {Craig}, {Dencheva}, {Ginsburg}, {Vand erPlas}, {Bradley}, {P{\'e}rez-Su{\'a}rez}, {de Val-Borro}, {Aldcroft}, {Cruz}, {Robitaille}, {Tollerud}, {Ardelean}, {Babej}, {Bach}, {Bachetti}, {Bakanov}, {Bamford}, {Barentsen}, {Barmby}, {Baumbach}, {Berry}, {Biscani}, {Boquien}, {Bostroem}, {Bouma}, {Brammer}, {Bray}, {Breytenbach}, {Buddelmeijer}, {Burke}, {Calderone}, {Cano Rodr{\'\i}guez}, {Cara}, {Cardoso}, {Cheedella}, {Copin}, {Corrales}, {Crichton}, {D'Avella}, {Deil}, {Depagne}, {Dietrich}, {Donath}, {Droettboom}, {Earl}, {Erben}, {Fabbro}, {Ferreira}, {Finethy}, {Fox}, {Garrison}, {Gibbons}, {Goldstein}, {Gommers}, {Greco}, {Greenfield}, {Groener}, {Grollier}, {Hagen}, {Hirst}, {Homeier}, {Horton}, {Hosseinzadeh}, {Hu}, {Hunkeler}, {Ivezi{\'c}}, {Jain}, {Jenness}, {Kanarek}, {Kendrew}, {Kern}, {Kerzendorf}, {Khvalko}, {King}, {Kirkby}, {Kulkarni},
  {Kumar}, {Lee}, {Lenz}, {Littlefair}, {Ma}, {Macleod}, {Mastropietro}, {McCully}, {Montagnac}, {Morris}, {Mueller}, {Mumford}, {Muna}, {Murphy}, {Nelson}, {Nguyen}, {Ninan}, {N{\"o}the}, {Ogaz}, {Oh}, {Parejko}, {Parley}, {Pascual}, {Patil}, {Patil}, {Plunkett}, {Prochaska}, {Rastogi}, {Reddy Janga}, {Sabater}, {Sakurikar}, {Seifert}, {Sherbert}, {Sherwood-Taylor}, {Shih}, {Sick}, {Silbiger}, {Singanamalla}, {Singer}, {Sladen}, {Sooley}, {Sornarajah}, {Streicher}, {Teuben}, {Thomas}, {Tremblay}, {Turner}, {Terr{\'o}n}, {van Kerkwijk}, {de la Vega}, {Watkins}, {Weaver}, {Whitmore}, {Woillez}, {Zabalza}, \& {Astropy Contributors}}]{astropy:2018}
{Astropy Collaboration}, {Price-Whelan}, A.~M., {Sip{\H{o}}cz}, B.~M., {et~al.} 2018, \aj, 156, 123, \dodoi{10.3847/1538-3881/aabc4f}

\bibitem[{{Astropy Collaboration} {et~al.}(2022){Astropy Collaboration}, {Price-Whelan}, {Lim}, {Earl}, {Starkman}, {Bradley}, {Shupe}, {Patil}, {Corrales}, {Brasseur}, {N{"o}the}, {Donath}, {Tollerud}, {Morris}, {Ginsburg}, {Vaher}, {Weaver}, {Tocknell}, {Jamieson}, {van Kerkwijk}, {Robitaille}, {Merry}, {Bachetti}, {G{"u}nther}, {Aldcroft}, {Alvarado-Montes}, {Archibald}, {B{'o}di}, {Bapat}, {Barentsen}, {Baz{'a}n}, {Biswas}, {Boquien}, {Burke}, {Cara}, {Cara}, {Conroy}, {Conseil}, {Craig}, {Cross}, {Cruz}, {D'Eugenio}, {Dencheva}, {Devillepoix}, {Dietrich}, {Eigenbrot}, {Erben}, {Ferreira}, {Foreman-Mackey}, {Fox}, {Freij}, {Garg}, {Geda}, {Glattly}, {Gondhalekar}, {Gordon}, {Grant}, {Greenfield}, {Groener}, {Guest}, {Gurovich}, {Handberg}, {Hart}, {Hatfield-Dodds}, {Homeier}, {Hosseinzadeh}, {Jenness}, {Jones}, {Joseph}, {Kalmbach}, {Karamehmetoglu}, {Ka{l}uszy{'n}ski}, {Kelley}, {Kern}, {Kerzendorf}, {Koch}, {Kulumani}, {Lee}, {Ly}, {Ma}, {MacBride}, {Maljaars}, {Muna}, {Murphy}, {Norman}, {O'Steen},
  {Oman}, {Pacifici}, {Pascual}, {Pascual-Granado}, {Patil}, {Perren}, {Pickering}, {Rastogi}, {Roulston}, {Ryan}, {Rykoff}, {Sabater}, {Sakurikar}, {Salgado}, {Sanghi}, {Saunders}, {Savchenko}, {Schwardt}, {Seifert-Eckert}, {Shih}, {Jain}, {Shukla}, {Sick}, {Simpson}, {Singanamalla}, {Singer}, {Singhal}, {Sinha}, {Sip{H{o}}cz}, {Spitler}, {Stansby}, {Streicher}, {{{S}}umak}, {Swinbank}, {Taranu}, {Tewary}, {Tremblay}, {Val-Borro}, {Van Kooten}, {Vasovi{'c}}, {Verma}, {de Miranda Cardoso}, {Williams}, {Wilson}, {Winkel}, {Wood-Vasey}, {Xue}, {Yoachim}, {Zhang}, {Zonca}, \& {Astropy Project Contributors}}]{astropy:2022}
{Astropy Collaboration}, {Price-Whelan}, A.~M., {Lim}, P.~L., {et~al.} 2022, \apj, 935, 167, \dodoi{10.3847/1538-4357/ac7c74}

\bibitem[{{Ballesteros-Paredes} {et~al.}(2011){Ballesteros-Paredes}, {Hartmann}, {V{\'a}zquez-Semadeni}, {Heitsch}, \& {Zamora-Avil{\'e}s}}]{Ballesteros-Paredes2011}
{Ballesteros-Paredes}, J., {Hartmann}, L.~W., {V{\'a}zquez-Semadeni}, E., {Heitsch}, F., \& {Zamora-Avil{\'e}s}, M.~A. 2011, \mnras, 411, 65, \dodoi{10.1111/j.1365-2966.2010.17657.x}

\bibitem[{{Beaumont} {et~al.}(2015){Beaumont}, {Goodman}, \& {Greenfield}}]{glue2015}
{Beaumont}, C., {Goodman}, A., \& {Greenfield}, P. 2015, in Astronomical Society of the Pacific Conference Series, Vol. 495, Astronomical Data Analysis Software an Systems XXIV (ADASS XXIV), ed. A.~R. {Taylor} \& E.~{Rosolowsky}, 101

\bibitem[{{Benedettini} {et~al.}(2020){Benedettini}, {Molinari}, {Baldeschi}, {Beltr{\'a}n}, {Brand}, {Cesaroni}, {Elia}, {Fontani}, {Merello}, {Olmi}, {Pezzuto}, {Rygl}, {Schisano}, {Testi}, \& {Traficante}}]{Benedettini2020}
{Benedettini}, M., {Molinari}, S., {Baldeschi}, A., {et~al.} 2020, \aap, 633, A147, \dodoi{10.1051/0004-6361/201936096}

\bibitem[{{Bolatto} {et~al.}(2013){Bolatto}, {Wolfire}, \& {Leroy}}]{Bolatto2013}
{Bolatto}, A.~D., {Wolfire}, M., \& {Leroy}, A.~K. 2013, \araa, 51, 207, \dodoi{10.1146/annurev-astro-082812-140944}

\bibitem[{{Brunt} \& {Mac Low}(2004)}]{Brunt2004}
{Brunt}, C.~M., \& {Mac Low}, M.-M. 2004, \apj, 604, 196, \dodoi{10.1086/381648}

\bibitem[{{Cen}(2021)}]{Cen2021}
{Cen}, R. 2021, \apjl, 906, L4, \dodoi{10.3847/2041-8213/abcecb}

\bibitem[{{Chevance} {et~al.}(2023){Chevance}, {Krumholz}, {McLeod}, {Ostriker}, {Rosolowsky}, \& {Sternberg}}]{Chevance2023}
{Chevance}, M., {Krumholz}, M.~R., {McLeod}, A.~F., {et~al.} 2023, in Astronomical Society of the Pacific Conference Series, Vol. 534, Protostars and Planets VII, ed. S.~{Inutsuka}, Y.~{Aikawa}, T.~{Muto}, K.~{Tomida}, \& M.~{Tamura}, 1, \dodoi{10.48550/arXiv.2203.09570}

\bibitem[{{Colman} {et~al.}(2024){Colman}, {Brucy}, {Girichidis}, {Glover}, {Benedettini}, {Soler}, {Tress}, {Traficante}, {Hennebelle}, {Klessen}, {Molinari}, \& {Miville-Desch{\^e}nes}}]{Colman2024}
{Colman}, T., {Brucy}, N., {Girichidis}, P., {et~al.} 2024, \aap, 686, A155, \dodoi{10.1051/0004-6361/202348983}

\bibitem[{{Dame} {et~al.}(2001){Dame}, {Hartmann}, \& {Thaddeus}}]{Dame2001}
{Dame}, T.~M., {Hartmann}, D., \& {Thaddeus}, P. 2001, \apj, 547, 792, \dodoi{10.1086/318388}

\bibitem[{{Dobbs} {et~al.}(2011){Dobbs}, {Burkert}, \& {Pringle}}]{Dobbs2011}
{Dobbs}, C.~L., {Burkert}, A., \& {Pringle}, J.~E. 2011, \mnras, 413, 2935, \dodoi{10.1111/j.1365-2966.2011.18371.x}

\bibitem[{{Elmegreen} \& {Scalo}(2004)}]{Elmegreen2004}
{Elmegreen}, B.~G., \& {Scalo}, J. 2004, \araa, 42, 211, \dodoi{10.1146/annurev.astro.41.011802.094859}

\bibitem[{{Fleck}(1980)}]{Fleck1980}
{Fleck}, R.~C., J. 1980, \apj, 242, 1019, \dodoi{10.1086/158533}

\bibitem[{{Fleck}(1981)}]{Fleck1981}
---. 1981, \apjl, 246, L151, \dodoi{10.1086/183573}

\bibitem[{{Garc{\'\i}a} {et~al.}(2014){Garc{\'\i}a}, {Bronfman}, {Nyman}, {Dame}, \& {Luna}}]{Garcia2014}
{Garc{\'\i}a}, P., {Bronfman}, L., {Nyman}, L.-{\r{A}}., {Dame}, T.~M., \& {Luna}, A. 2014, \apjs, 212, 2, \dodoi{10.1088/0067-0049/212/1/2}

\bibitem[{{Grisdale} {et~al.}(2018){Grisdale}, {Agertz}, {Renaud}, \& {Romeo}}]{Grisdale2018}
{Grisdale}, K., {Agertz}, O., {Renaud}, F., \& {Romeo}, A.~B. 2018, \mnras, 479, 3167, \dodoi{10.1093/mnras/sty1595}

\bibitem[{Harris {et~al.}(2020)Harris, Millman, van~der Walt, Gommers, Virtanen, Cournapeau, Wieser, Taylor, Berg, Smith, Kern, Picus, Hoyer, van Kerkwijk, Brett, Haldane, del R{\'{i}}o, Wiebe, Peterson, G{\'{e}}rard-Marchant, Sheppard, Reddy, Weckesser, Abbasi, Gohlke, \& Oliphant}]{numpy2020}
Harris, C.~R., Millman, K.~J., van~der Walt, S.~J., {et~al.} 2020, Nature, 585, 357, \dodoi{10.1038/s41586-020-2649-2}

\bibitem[{{Heyer} {et~al.}(2009){Heyer}, {Krawczyk}, {Duval}, \& {Jackson}}]{Heyer2009}
{Heyer}, M., {Krawczyk}, C., {Duval}, J., \& {Jackson}, J.~M. 2009, \apj, 699, 1092, \dodoi{10.1088/0004-637X/699/2/1092}

\bibitem[{{Heyer} {et~al.}(2001){Heyer}, {Carpenter}, \& {Snell}}]{Heyer2001}
{Heyer}, M.~H., {Carpenter}, J.~M., \& {Snell}, R.~L. 2001, \apj, 551, 852, \dodoi{10.1086/320218}

\bibitem[{{Hughes} {et~al.}(2013){Hughes}, {Meidt}, {Colombo}, {Schinnerer}, {Pety}, {Leroy}, {Dobbs}, {Garc{\'\i}a-Burillo}, {Thompson}, {Dumas}, {Schuster}, \& {Kramer}}]{Hughes2013}
{Hughes}, A., {Meidt}, S.~E., {Colombo}, D., {et~al.} 2013, \apj, 779, 46, \dodoi{10.1088/0004-637X/779/1/46}

\bibitem[{{Ib{\'a}{\~n}ez-Mej{\'\i}a} {et~al.}(2016){Ib{\'a}{\~n}ez-Mej{\'\i}a}, {Mac Low}, {Klessen}, \& {Baczynski}}]{IBANEZ2016}
{Ib{\'a}{\~n}ez-Mej{\'\i}a}, J.~C., {Mac Low}, M.-M., {Klessen}, R.~S., \& {Baczynski}, C. 2016, \apj, 824, 41, \dodoi{10.3847/0004-637X/824/1/41}

\bibitem[{{Ib{\'a}{\~n}ez-Mej{\'\i}a} {et~al.}(2017){Ib{\'a}{\~n}ez-Mej{\'\i}a}, {Mac Low}, {Klessen}, \& {Baczynski}}]{Ibanez2017}
---. 2017, \apj, 850, 62, \dodoi{10.3847/1538-4357/aa93fe}

\bibitem[{{Kauffmann} {et~al.}(2013){Kauffmann}, {Pillai}, \& {Goldsmith}}]{Kauffmann2013}
{Kauffmann}, J., {Pillai}, T., \& {Goldsmith}, P.~F. 2013, \apj, 779, 185, \dodoi{10.1088/0004-637X/779/2/185}

\bibitem[{{Kauffmann} {et~al.}(2017){Kauffmann}, {Pillai}, {Zhang}, {Menten}, {Goldsmith}, {Lu}, \& {Guzm{\'a}n}}]{Kauffmann2017}
{Kauffmann}, J., {Pillai}, T., {Zhang}, Q., {et~al.} 2017, \aap, 603, A89, \dodoi{10.1051/0004-6361/201628088}

\bibitem[{{Lada} \& {Dame}(2020)}]{Lada2020}
{Lada}, C.~J., \& {Dame}, T.~M. 2020, \apj, 898, 3, \dodoi{10.3847/1538-4357/ab9bfb}

\bibitem[{{Larson}(1981)}]{Larson1981}
{Larson}, R.~B. 1981, \mnras, 194, 809, \dodoi{10.1093/mnras/194.4.809}

\bibitem[{{Li} {et~al.}(2019){Li}, {Zhao}, \& {Yang}}]{Li2019oort}
{Li}, C., {Zhao}, G., \& {Yang}, C. 2019, \apj, 872, 205, \dodoi{10.3847/1538-4357/ab0104}

\bibitem[{{Liu} {et~al.}(2021){Liu}, {Bureau}, {Blitz}, {Davis}, {Onishi}, {Smith}, {North}, \& {Iguchi}}]{Liu2021}
{Liu}, L., {Bureau}, M., {Blitz}, L., {et~al.} 2021, \mnras, 505, 4048, \dodoi{10.1093/mnras/stab1537}

\bibitem[{{Luo} {et~al.}(2024){Luo}, {Liu}, {Qin}, {Yang}, \& {Pan}}]{Luo2024}
{Luo}, A.-X., {Liu}, H.-L., {Qin}, S.-L., {Yang}, D.-t., \& {Pan}, S. 2024, \aj, 167, 228, \dodoi{10.3847/1538-3881/ad35ca}

\bibitem[{{McKee} \& {Ostriker}(2007)}]{McKee2007}
{McKee}, C.~F., \& {Ostriker}, E.~C. 2007, \araa, 45, 565, \dodoi{10.1146/annurev.astro.45.051806.110602}

\bibitem[{{McKee} \& {Zweibel}(1992)}]{McKee1992}
{McKee}, C.~F., \& {Zweibel}, E.~G. 1992, \apj, 399, 551, \dodoi{10.1086/171946}

\bibitem[{{Miville-Desch{\^e}nes} {et~al.}(2017){Miville-Desch{\^e}nes}, {Murray}, \& {Lee}}]{Miville2017}
{Miville-Desch{\^e}nes}, M.-A., {Murray}, N., \& {Lee}, E.~J. 2017, \apj, 834, 57, \dodoi{10.3847/1538-4357/834/1/57}

\bibitem[{{Mr{\'o}z} {et~al.}(2019){Mr{\'o}z}, {Udalski}, {Skowron}, {Skowron}, {Soszy{\'n}ski}, {Pietrukowicz}, {Szyma{\'n}ski}, {Poleski}, {Koz{\l}owski}, \& {Ulaczyk}}]{Mroz2019}
{Mr{\'o}z}, P., {Udalski}, A., {Skowron}, D.~M., {et~al.} 2019, \apjl, 870, L10, \dodoi{10.3847/2041-8213/aaf73f}

\bibitem[{{Nemmen} {et~al.}(2012){Nemmen}, {Georganopoulos}, {Guiriec}, {Meyer}, {Gehrels}, \& {Sambruna}}]{Nemmen2012}
{Nemmen}, R.~S., {Georganopoulos}, M., {Guiriec}, S., {et~al.} 2012, Science, 338, 1445, \dodoi{10.1126/science.1227416}

\bibitem[{{Rice} {et~al.}(2016){Rice}, {Goodman}, {Bergin}, {Beaumont}, \& {Dame}}]{Rice2016}
{Rice}, T.~S., {Goodman}, A.~A., {Bergin}, E.~A., {Beaumont}, C., \& {Dame}, T.~M. 2016, \apj, 822, 52, \dodoi{10.3847/0004-637X/822/1/52}

\bibitem[{{Robitaille} {et~al.}(2017){Robitaille}, {Beaumont}, {Qian}, {Borkin}, \& {Goodman}}]{glue2017}
{Robitaille}, T., {Beaumont}, C., {Qian}, P., {Borkin}, M., \& {Goodman}, A. 2017, {glueviz v0.13.1: multidimensional data exploration}, 0.13.1,  Zenodo, \dodoi{10.5281/zenodo.1237692}

\bibitem[{{Rosolowsky} {et~al.}(2008){Rosolowsky}, {Pineda}, {Kauffmann}, \& {Goodman}}]{Rosolowsky2008}
{Rosolowsky}, E.~W., {Pineda}, J.~E., {Kauffmann}, J., \& {Goodman}, A.~A. 2008, \apj, 679, 1338, \dodoi{10.1086/587685}

\bibitem[{{Shetty} {et~al.}(2012){Shetty}, {Beaumont}, {Burton}, {Kelly}, \& {Klessen}}]{Shetty2012}
{Shetty}, R., {Beaumont}, C.~N., {Burton}, M.~G., {Kelly}, B.~C., \& {Klessen}, R.~S. 2012, \mnras, 425, 720, \dodoi{10.1111/j.1365-2966.2012.21588.x}

\bibitem[{{Solomon} {et~al.}(1987){Solomon}, {Rivolo}, {Barrett}, \& {Yahil}}]{Solomon1987}
{Solomon}, P.~M., {Rivolo}, A.~R., {Barrett}, J., \& {Yahil}, A. 1987, \apj, 319, 730, \dodoi{10.1086/165493}

\bibitem[{{Sun} {et~al.}(2024){Sun}, {Yang}, {Yan}, {Zhang}, {Su}, {Chen}, {Zhou}, {Ma}, \& {Yuan}}]{Sun2024}
{Sun}, Y., {Yang}, J., {Yan}, Q.-Z., {et~al.} 2024, \apjs, 275, 35, \dodoi{10.3847/1538-4365/ad8237}

\bibitem[{{Traficante} {et~al.}(2018){Traficante}, {Duarte-Cabral}, {Elia}, {Fuller}, {Merello}, {Molinari}, {Peretto}, {Schisano}, \& {Di Giorgio}}]{Traficante2018}
{Traficante}, A., {Duarte-Cabral}, A., {Elia}, D., {et~al.} 2018, \mnras, 477, 2220, \dodoi{10.1093/mnras/sty798}

\bibitem[{{Vergely} {et~al.}(2022){Vergely}, {Lallement}, \& {Cox}}]{Vergely2022}
{Vergely}, J.~L., {Lallement}, R., \& {Cox}, N.~L.~J. 2022, \aap, 664, A174, \dodoi{10.1051/0004-6361/202243319}

\bibitem[{Virtanen {et~al.}(2020)Virtanen, Gommers, Oliphant, Haberland, Reddy, Cournapeau, Burovski, Peterson, Weckesser, Bright, {van der Walt}, Brett, Wilson, Millman, Mayorov, Nelson, Jones, Kern, Larson, Carey, Polat, Feng, Moore, {VanderPlas}, Laxalde, Perktold, Cimrman, Henriksen, Quintero, Harris, Archibald, Ribeiro, Pedregosa, {van Mulbregt}, \& {SciPy 1.0 Contributors}}]{scipy2020}
Virtanen, P., Gommers, R., Oliphant, T.~E., {et~al.} 2020, Nature Methods, 17, 261, \dodoi{10.1038/s41592-019-0686-2}

\bibitem[{{Wada} {et~al.}(2002){Wada}, {Meurer}, \& {Norman}}]{Wada2002}
{Wada}, K., {Meurer}, G., \& {Norman}, C.~A. 2002, \apj, 577, 197, \dodoi{10.1086/342151}

\bibitem[{{Xie} {et~al.}(2024){Xie}, {Li}, \& {Chen}}]{Xie2024}
{Xie}, Y.-H., {Li}, G.-X., \& {Chen}, B.-Q. 2024, \apj, 975, 39, \dodoi{10.3847/1538-4357/ad7378}

\bibitem[{{Zhou} {et~al.}(2022){Zhou}, {Li}, \& {Chen}}]{zhou2022}
{Zhou}, J.-X., {Li}, G.-X., \& {Chen}, B.-Q. 2022, \mnras, 513, 638, \dodoi{10.1093/mnras/stac900}

\end{thebibliography}
\bibliographystyle{aasjournal}

\end{document}